\newif\ifwordcount
\wordcountfalse

\documentclass[%
 reprint,
superscriptaddress,
showpacs,
nofootinbib,
 amsmath,amssymb,
 amsart,
 aps,
prl,
tightenlines,
floatfix,
twocolumn
]{revtex4-2}

\usepackage{graphicx}% Include figure files
\usepackage{dcolumn}% Align table columns on decimal point
\usepackage{bm}% bold math
\usepackage{color}
\definecolor{linkcolor}{rgb}{0.6,0,0}
\definecolor{citecolor}{rgb}{0,0,0.75}
\definecolor{urlcolor}{rgb}{0.12,0.46,0.7}
\usepackage[dvipsnames]{xcolor}

\usepackage[breaklinks, colorlinks, urlcolor=urlcolor, linkcolor=linkcolor,citecolor=citecolor,pdfencoding=auto]{hyperref}
\usepackage{amsmath}
\usepackage{xspace}
\usepackage{booktabs}
\usepackage{siunitx}
\usepackage{xcolor,colortbl}
\usepackage[english]{babel}
\usepackage{comment}

\newcommand{\planck}{\textit{Planck}\xspace}
\newcommand{\aap}{Astron. Astrophys.}
\newcommand{\mnras}{Mon. Not. Roy. Astron. Soc.}

\newcommand{\jcap}{JCAP}
\newcommand{\pasp}{Publ. Astron. Soc. Pac.}

\newcommand{\nver}{\hat{\mathbf{n}}}

\newcommand{\ymono}{\langle y\rangle}
\newcommand{\tmono}{\langle T_e\rangle}
\newcommand{\msun}{M_{\odot}\xspace}

\begin{document}

\defcitealias{sabyr2025}{S25}
\defcitealias{Abitbol2017}{A17}
\defcitealias{BianchiniFabbian2022}{BF22}
\defcitealias{ourmu2022}{BF22}
\defcitealias{Abitbol2017}{A17}
\defcitealias{fixsen96}{F96}
\defcitealias{thiele2022}{T22}
\defcitealias{fes}{FES}

\newcommand{\ourmu}{\citetalias{ourmu2022}\xspace}
\newcommand{\sabyr}{\citetalias{sabyr2025}\xspace}
\newcommand{\thiele}{\citetalias{thiele2022}\xspace}
\let\sectionoold\section
\preprint{}

\title{A new constraint on the $y$-distortion with FIRAS: implications for feedback models in galaxy formation and cosmic shear measurements}

\author{Giulio Fabbian}%
\email{giulio.fabbian@universite-paris-saclay.fr}
\affiliation{Université Paris-Saclay, CNRS, Institut d’Astrophysique Spatiale, 91405, Orsay, France}
\affiliation{School of Physics and Astronomy, Cardiff University, The Parade, Cardiff, CF24 3AA, UK}
\affiliation{Kavli Institute for Cosmology Cambridge, Madingley Road, Cambridge CB3 0HA, UK}

\author{Federico Bianchini}%
\email{fbianc@stanford.edu}
\affiliation{Kavli Institute for Particle Astrophysics and Cosmology, Stanford University, 452 Lomita Mall, Stanford, CA, 94305, USA}
\affiliation{SLAC National Accelerator Laboratory, 2575 Sand Hill Road, Menlo Park, CA, 94025, USA}
\affiliation{Department of Physics, Stanford University, 382 Via Pueblo Mall, Stanford, CA, 94305, USA}

\author{Alina Sabyr}
\affiliation{Berkeley Center for Cosmological Physics, Department of Physics, University of California, Berkeley, CA 94720, USA}
\affiliation{Lawrence Berkeley National Laboratory, One Cyclotron Road, Berkeley, CA 94720, USA}
\affiliation{Department of Astronomy, Columbia University New York, New York, USA 10027}

\author{J.~Colin Hill}
\affiliation{Department of Physics, Columbia University New York, NY, USA 10027}

\author{Christopher C. Lovell}
\affiliation{Kavli Institute for Cosmology Cambridge, Madingley Road, Cambridge CB3 0HA, UK}
\affiliation{Institute of Astronomy, Madingley Road, Cambridge CB3 0HA, GB}

\author{Leander Thiele}%
\affiliation{Kavli Institute for the Physics and Mathematics of the Universe (WPI), The University of Tokyo Institutes for Advanced Study (UTIAS), The University of Tokyo, Chiba 277-8583, Japan}
\affiliation{Center for Data-Driven Discovery, Kavli IPMU (WPI), UTIAS, The University of Tokyo, Kashiwa, Chiba 277-8583, Japan}

\author{David N. Spergel}
\affiliation{Center for Computational Astrophysics, Flatiron Institute, New York, NY 10010, USA}
\affiliation{Department of Astrophysical Sciences, Princeton University, 4 Ivy Ln, Princeton, NJ 08544, USA}

\date{\today}

\begin{abstract}
The $y$-type distortion of the blackbody spectrum of the cosmic microwave background radiation probes the pressure of the gas trapped in galaxy groups and clusters. We reanalyze archival data of the FIRAS instrument with an improved astrophysical foreground cleaning technique, and measure a mean $y$-distortion of $\langle y\rangle = (1.2\pm 2.0) \times 10^{-6}$ ($\langle y\rangle\lesssim 5.2\times 10^{-6}$ at 95\% C.L.), a factor of $\sim 3$ tighter than the original FIRAS results. This measurement directly rules out many models of baryonic feedback as implemented in cosmological hydrodynamical simulations, mostly using information in objects with mass $M\lesssim 10^{14}\msun$. We discuss its implications for the analysis of cosmic shear and kinetic Sunyaev-Zel'dovich effect data, and future spectral distortion experiments.
\end{abstract}

\ifwordcount
\else
\maketitle
\fi

\section{Introduction}
The blackbody nature of the spectral energy distribution (SED) of the cosmic microwave background (CMB) radiation is one of the key predictions of the hot Big Bang cosmology. Its observation in the early 1990s provided an essential observable to understand the thermal history of the Universe and stands as one of the cornerstones of the current cosmological model \citep{mather93,zannoni08,fixsen11}. Departures from the perfect blackbody SED (spectral distortions)  arise when thermalization processes are inefficient in keeping matter and radiation in thermodynamical equilibrium, and can be used to constrain a wide variety of physical processes acting before and after the recombination era \citep[][and references therein]{chluba19}. 
While the so-called $\mu$-type distortion \citep[][]{sz70,burigana91} probes the physics of the early universe ($z\gtrsim 5\times 10^4$) \citep{pajer12,ganc12,Biagetti:2013sr,emami15,ota16,ravenni17,Cabass:2018jgj,Lucca2019,Fu2020,orlando21},  the $y$-type distortion is sourced at $z\lesssim 5\times 10^4$ when Compton scattering becomes inefficient and photons fall out of kinetic equilibrium with electrons. Thus, $y$ distortions can be generated both in the early \cite{Bolliet:2020ofj} and late Universe, since any energy output from the first stars, accreting black holes, and gravitational shocks can all heat the baryons and electrons. These can then scatter off CMB photons producing $y$ distortions \citep[][]{zeldovich72, hu94,refregier00}. \\
The expected average $y$ distortion across the sky, $\ymono$, is $\mathcal{O}(10^{-6})$ \cite{hill2015,barbosa1996,Dolag2016}. The leading contribution to the signal comes from the inverse-Compton scattering of photons off electrons residing in virialized dark matter halos with masses $M \approx 10^{12.5} - 10^{14.5} \, \msun$ and temperatures $k_BT_e \approx 2 - 3$~keV at $z\lesssim 2$ (thermal Sunyaev-Zel’dovich effect, tSZ \citep{zeldovich69,carlstrom02,mroczkowski19}). This can be written as \cite{hill2015}

\begin{equation}
\ymono = \frac{\sigma_T}{m_ec^2}\int \frac{d^{2}\hat{\mathbf{n}}}{4\pi}\int dl P_e(\hat{\mathbf{n}},l) \propto E^{\rm th,total}_e
\end{equation}
where $P_e\equiv n_ek_{\rm B}T_e$ is the electron pressure (with $n_e$ the electron number density), $l$ the line of sight, $\sigma_T$ the Thomson cross-section, $m_ec^2$ the electron rest-mass energy, and $E^{\rm th,total}_e$ the total thermal energy in electrons including contributions from gravitational collapse, cooling, and other energy injections.  As such, $\ymono$ is a unique probe to test the thermodynamical processes affecting hot electron gas in such environments, as well as the abundance of dark matter halos (the halo mass function), which ultimately depends on the cosmological model.  
Relativistic corrections to the tSZ-induced spectral shape of the $y$ distortion (relevant when $k_BT_e/m_ec^2\gtrsim 10^{-2}$) can also directly provide insights into $T_e$, which at present is constrained by X-ray observations with large uncertainties \cite{remazeilles-rsz,remazeilles2024,coulton2024}.\\ 

The Far Infrared Absolute Spectrophotometer (FIRAS) instrument of the \textit{COBE} satellite \cite{mather93} set the first robust observational bounds on $\ymono$: $|\langle y\rangle|<15 \times 10^{-6}$ (95\% C.L.) ($\ymono = (-1\pm 6\ {\rm stat. } \pm 4\ {\rm syst.})\times 10^{-6}$) in the 1990s \citep{fixsen96}. 
Several follow-up experiments or instrumental concepts (e.g., PIXIE, PRISM, BISOU, COSMO, SPECTER, FOSSIL \citep{prism, pixie2016, bisou, cosmo, specter, fossil}) have been proposed to improve the FIRAS measurements, but to date no new data have been acquired. 
In this letter we reanalyze FIRAS data with an improved astrophysical foreground cleaning approach, and derive the tightest constraints on $\ymono$ to date. This work complements \cite[][]{ourmu2022} (hereafter \citetalias{ourmu2022}), where we improved the constraints on $\mu$ distortions to $|\langle \mu\rangle|< 47\times 10^{-6}$ (95\%. C.L.) and \cite[][]{sabyr2025} (hereafter \citetalias{sabyr2025}), which investigated foreground cleaning strategies for $\ymono$ measurements. In particular, we investigate the implications of the more constraining method proposed therein. 

\section{\label{sec:data}Data sets} 
FIRAS was a cryogenically cooled Martin-Puplett interferometer that provided absolutely calibrated full-sky brightness measurements with $\sim 7^\circ$ angular resolution. For this work we use its final data release, reprocessed in HEALPix pixelization and corrected for systematics and low-frequency noise through a destriping procedure.  These maps are publicly available on the NASA LAMBDA archive\footnote{\url{https://lambda.gsfc.nasa.gov/product/cobe/firas_tpp_all_get.cfm}}. A detailed discussion of the instrument, its data analysis, and data products can be found in the FIRAS explanatory supplement (FES) \cite{fes}.  
We make use of Galactic masks constructed for the \planck data analysis \cite{planck18_likelihood}, which remove pixels with high Galactic foreground contamination. These retain sky fractions ranging from 20\% to 90\%\footnote{See the \href{http://pla.esac.esa.int/pla/\#home}{Planck legacy archive}.}. We denote these as, e.g., P20 or P90, respectively. To all these masks we add the so-called FIRAS destriper mask that defines the FIRAS sky coverage and removes pixels with large systematic uncertainties. 

\section{Data model}\label{sec:methods} 
Following \citetalias{ourmu2022,sabyr2025}, we use a pixel-by-pixel component separation method to isolate the $y$ distortion from astrophysical foreground emission. The FIRAS data are treated as pixelized maps of the sky’s absolute brightness at multiple frequencies. At each frequency $\nu$, the sky emission in direction $\nver$ is modeled as the sum of the CMB blackbody spectrum $B_{\nu}(T_0)$ at $T_0=2.7255$~K, brightness variations $\Delta T(\nver)$ due to CMB temperature anisotropies, foreground emission $I_{\nu}^{\rm FG}(\nver)$ described by a set of parameters, and the $y$ distortion:

\ifwordcount
\else
\begin{equation}
    I_{\nu}(\nver) = B_{\nu}(T_0) + \Delta T(\nver) \frac{\partial B_{\nu}}{\partial T}\bigg\rvert_{T_0} +I_{\nu}^{\rm FG}(\nver) + y(\nver)I_{\nu}^{y}.
\label{eq:data-model}
\end{equation}
\fi
where the explicit frequency dependence of $I_\nu^y$ is given in \cite{sz69}.  We then fit the sky model parameters $\bm\theta$  ($y, \Delta T$, foreground parameters) in each individual sky pixel using a likelihood: 

\ifwordcount
\else
\begin{equation}\label{eq:pix_like}
    -2\ln\mathcal{L}(\hat{I}_{\nu}|\bm\theta) \propto \sum_{\nu\nu'}\Delta_{\nu}^{T}(\bm\theta) \mathbb{C}^{-1}_{\nu\nu'} \Delta_{\nu'}(\bm\theta),
\end{equation}
\fi
where $\Delta_{\nu}(\bm\theta) = \hat{I}_{\nu}-I_{\nu}(\bm\theta)$ denotes the residuals between the observed FIRAS spectra and the sky model, and $\mathbb{C}^{}_{\nu\nu'}$ is the FIRAS full frequency-frequency covariance matrix in that specific pixel.  An affine-invariant MCMC sampler \citep{emcee} is used to explore the posterior distributions and produce maps of $y$, $\Delta T$, foreground parameters, and their uncertainties. The mean distortion $\ymono$ is then obtained by averaging the pixel values of the $y$ map using inverse-covariance weighting. 
We fit the sky model using frequencies from 60 to 600~GHz, divided into linearly spaced bins of $\Delta \nu\approx 13$~GHz. This range, shown by \citetalias{sabyr2025} to best constrain the $y$ distortion (whose SED has a negative/positive peak near 129/370~GHz), avoids complications at higher frequencies due to FIRAS calibration errors and increased foreground complexity. FIRAS data are correlated across both frequency $\nu$ and sky pixel $p$ due to instrumental effects and the destriping procedure, forming a full covariance matrix $\mathbb{C}_{\nu p p^\prime\nu^\prime}$\footnote{$\mathbb{C}_{\nu\nu'} $ in Eq.~\eqref{eq:pix_like} takes the $p=p^\prime$ part of this matrix retaining all the error terms described in the \citetalias{fes}.}. For simplicity, we use only the $p = p^\prime$ terms, retaining frequency–frequency correlations while neglecting inter-pixel ones, to reduce computational cost. Following \citetalias{sabyr2025}, to correct for this we estimate pixel–pixel correlations directly from the $y$ values in the MCMC chains and use them to build the covariance matrix used in the $\ymono$ calculation. This approach improves the inaccuracy of the original FIRAS correlation estimates \cite{sabyr2025}.

\section{Foregrounds}\label{sec:foregrounds}
The original FIRAS analysis used the $\sim\!1.2$~THz and $\sim\!2.1$~THz maps of the DIRBE instrument onboard \textit{COBE} as foreground templates to be adjusted linearly and subtracted from the data at each frequency before fitting $\ymono$ on the sky-averaged frequency spectrum \cite{fixsen96}. While being the only viable approach at the time, our FIRAS-based \textit{pixel-by-pixel} approach improves on this by accounting for the spatial variations in the foreground properties, and avoiding uncertainties from extrapolating foreground properties from $2$ THz to $\nu \ll 1$~THz. \citetalias{sabyr2025} also showed that fitting $\ymono$ directly from sky-averaged brightness measurements \cite[][]{Abitbol:2017vwa} is more susceptible to foreground residuals. 
Moreover, a recent reprocessing of DIRBE data revealed significant zodiacal light contamination in the original maps \cite{cosmoglobe2024dirbe,cosmoglobe2024zodi}. For these new DIRBE maps \cite{cosmoglobe-data} we find noise levels $0.17$~MJy/sr and $0.19$~MJy/sr for $\sim\!1.2$~THz and $\sim\!2.1$~THz respectively\footnote{We estimate this as the variance of the half-difference of maps derived from independent data halves.}, comparable to or higher than that of FIRAS in our frequency range, meaning their use would add correlated noise and reduce the sensitivity of the original analysis. 

The main astrophysical foregrounds in our analysis are Galactic thermal dust emission and the cosmic infrared background (CIB) produced by unresolved star-forming galaxies, which dominates at high Galactic latitudes. The CIB and tSZ signals are expected to be correlated at the 10-40\% level depending on frequency \cite{planck15_yXCIB}, as some dusty galaxies reside in galaxy groups and clusters~\cite{montier2005,giard2008,Addison2012,planck2016cibdust}. Their accurate separation is therefore crucial as CIB residuals can mimic a Compton-$y$ signal at the same location.
Following \citetalias{ourmu2022,sabyr2025}, we neglect synchrotron, anomalous microwave emission, free-free, and Galactic CO emission, as they are either negligible or absorbed by our foreground models \cite{so-galscience}. Unlike the conservative choices of \citetalias{sabyr2025}, we retain the lowest FIRAS frequency bands and those close to CO transition lines, verifying with mock data that this introduces no bias in $\ymono$. In addition, the $\ymono$ obtained here with this setup is consistent within $\sim 1.3\sigma$ with the results of \citetalias{sabyr2025} obtained with a similar \emph{pixel-by-pixel} approach.

\begin{figure}[t]
\includegraphics[width=\columnwidth]{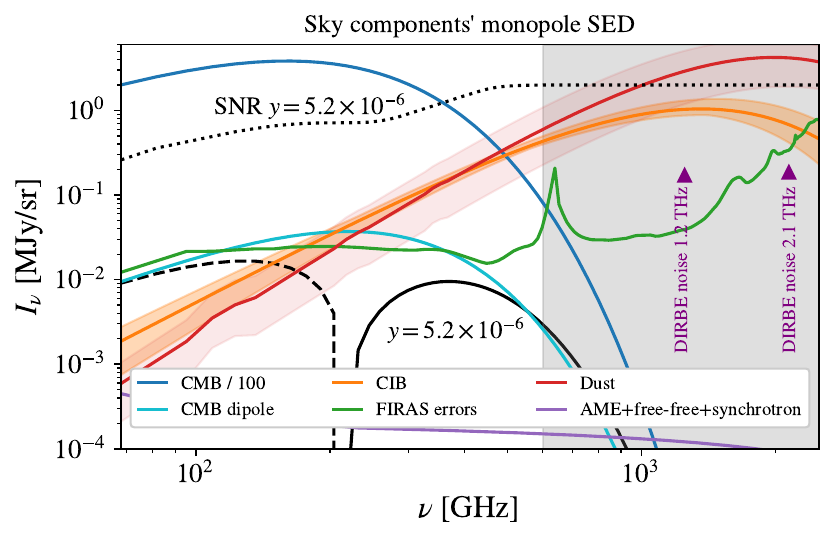}
\caption{SEDs of sky monopole components on the P60 footprint from the \textit{Planck} Commander sky model \cite{planck15_fg}, compared to the FIRAS total error (green). Our new $\ymono$ measurement and cumulative signal-to-noise ratio relative to FIRAS errors are in solid and dotted black. The grey area denotes excluded FIRAS frequencies. The shaded red shows the Galactic dust monopole on the P80 and P20 footprints. The orange line and shading represent the mean, maximum, and minimum CIB monopole amplitudes for our considered CIB SEDs. The purple points indicate foreground template noise from the original FIRAS analysis.}
\label{fig:all-sed}
\end{figure}

We model the Galactic dust SED with parameters $\{A_d, \beta_d, T_d\}$ as
\begin{equation}
    I_{\nu}^{\text {dust}}(\nver) = A_d(\nver) \left(\frac{\nu}{353 \, \rm GHz}\right)^{\beta_d(\nver)} \frac{B_{\nu}\left(T_{d}(\nver)\right)}{B_{353 \, \rm  GHz}\left(T_{d}(\nver)\right)}.
\label{eq:dust-model}
\end{equation}
and the CIB with amplitude $A_{\rm CIB}$ as 
\begin{equation}
    I_{\nu}^{\text {CIB }}(\nver) \propto A_{\rm CIB}(\nver) \nu^{\beta_{\rm CIB}} B_{\nu}\left(T_{\rm CIB}\right).
\label{eq:cib-model}
\end{equation}
where each SED is described by a modified blackbody  $B_\nu(T)\nu^\beta$ with temperature $T$ and spectral index $\beta$. Because dust and CIB have similar frequency scaling and overlap spatially, making their separation difficult \cite{planck15_fg}, we test several flexible combinations of free parameters to capture their superposition: $\{A_d , \beta_d\}, \{A_d ,\beta_d, A_{\rm CIB}\}, \{A_d, A_{\rm CIB}\}$, and $\{A_d ,\beta_d, T_d\}$.
When not varied, we either fix $T_d=19.6\,{\rm K}$ and/or $\beta_d=1.6$ to their mean values found by \planck \cite{planck13_dust,planck18_compsep}. For the CIB, we vary only $A_{\rm CIB}$ but adopt different fixed SEDs from \cite[][]{fixsen1998} (Fixsen, $\beta_{\rm CIB}=0.64, T_{\rm CIB}=18.8$~K), \cite[][]{gispert2000} (Gispert, $\beta_{\rm CIB}=1.4, T_{\rm CIB}=13.6$~K), and \cite[][]{Abitbol:2017vwa} (Abitbol, $\beta_{\rm CIB}=0.86, T_{\rm CIB}=18.8$~K). When varied, we adopt uniform priors $\beta_d\in [0,3]$, $T_d\in (0,100]$~K following \citetalias{sabyr2025}. These are sufficient to capture the superposition of dust and CIB emission, and tests on the same realistic mock data of \citetalias{sabyr2025} confirm that all the models we consider recover unbiased $\ymono$ values up to the P80 sky mask. Figure~\ref{fig:all-sed} shows the SEDs of all components considered in this work.

%\ifwordcount
%\else
%\fi

\section{Results}\label{sec:results}
To compute $\ymono$ from the estimated $y$ maps, we first build a confidence mask that retains only pixels where $y$ values are consistent across all component separation methods, reducing the impact of  foreground residuals and modeling uncertainties in our final result. Following \cite{gratton2019}, we test the consistency between two estimates ($a$ and $b$) of $y$ obtained with models with a nested set of parameters for each sky pixel $p$ as
\begin{equation}
\Delta y_p  = \frac{|y_{p,a} - y_{p,b}|}{\sqrt{ |\sigma_{y_{p,a}}^2 - \sigma_{y_{p,b}}^2 |}} \lesssim 2.0, % 2.5?
\label{eq:yshift}
\end{equation}
where $\sigma^2_{y_{p}}$ is the variance of the estimate obtained fitting model $a$ or $b$ to the data. Our total confidence mask is defined as the intersection of pixels consistent between  the $A_d+\beta_d+A_{\rm CIB}$ models and $A_d+\beta_d$ for all CIB SEDs, and between the $A_d+\beta_d$  and $A_d+\beta_d+T_d$ models\footnote{The confidence masks comparing $A_d+\beta_d+A_{\rm CIB}$ and $A_d+A_{\rm CIB}$ models for a fixed CIB SED do not remove additional pixels.}. Pixels outside this mask are excluded from further analysis.
We then compute $\ymono$ for each $y$ map on Galactic masks retaining progressively larger sky fractions (from P20 to P90). Because these masks are nested, we can apply the same criterion of Eq.~\eqref{eq:yshift} to check if the estimates obtained on two different footprints are consistent with statistical fluctuations or show inconsistent variations likely induced by foreground residuals \citep{gratton2019}. Because negative $y$ values often appear in regions with high foreground emission \cite{sabyr2025}, this approach helps avoiding a bias towards low $\ymono$.

\begin{figure}[htbp]
\includegraphics[width=\columnwidth]{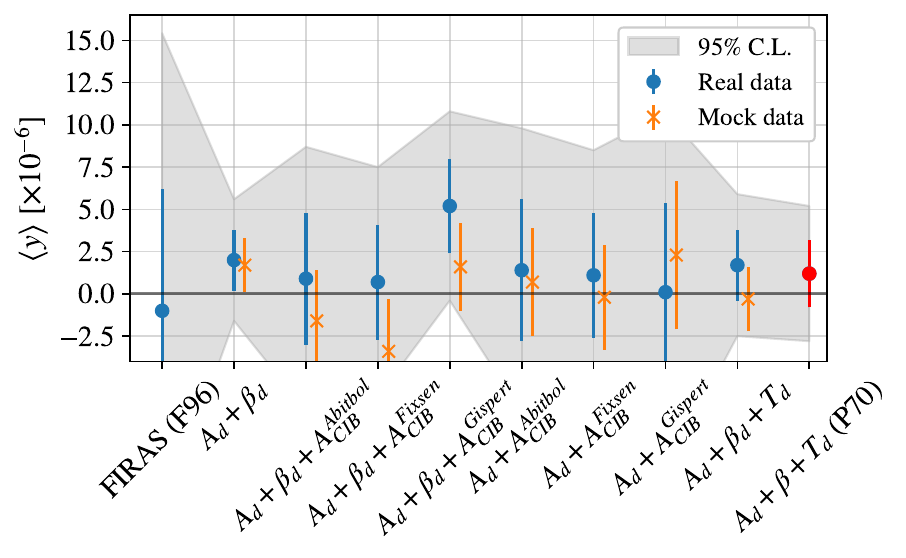}\\
\caption{$\ymono$ measurements obtained with different foreground cleaning methods on FIRAS real (blue) and mock (orange) data obtained on the P60 mask. Our best estimate is shown in red.}
\label{fig:ymono}
\end{figure}

Table~\ref{tab:ymono-allmasks} in the End Matter lists all $\ymono$ estimates for different foreground models and masks. Maps and results on mocks are provided in the supplemental material. 
We find that the P60 mask provides the largest sky fraction where all component separation methods yield consistent $\ymono$ estimates, within $2.1\sigma$ from those obtained on P40 and P20. In Figure~\ref{fig:ymono}  we show all these values  together with the previous \citetalias{fixsen96} results. All the component separation methods yield $\ymono$ consistent with null at 95\% C.L., with errors matching expectations from simulated data. Masking pixels where the best-fit $\chi^2$ has a $p$-value $<0.001$ does not affect our results. The $A_d+\beta_d$ foreground model delivers the tightest constraint, $\ymono = \left(2.0 \pm 1.8\right) \times 10^{-6}$ ($\ymono<5.5\times 10^{-6}$ at 95\% C.L.), but becomes unreliable at low Galactic latitudes.  Any additional sky area included by the P70, P80, or P90 masks leads to estimates inconsistent with those obtained on smaller sky fractions. In contrast, the $A_d+\beta_d+T_d$ and $A_d+\beta_d+A_{\rm CIB}$ models display minor inconsistencies, though with larger uncertainties. Among those, $A_d+\beta_d+T_d$ is the most general model as it does not introduce assumptions on the CIB SED, and its two non-linear degrees of freedom introduce a lower loss of sensitivity for the addition of an extra foreground parameter compared to $A_d+\beta_d$ (and contrary to $A_d+\beta_d+A_{\rm CIB}$). Additional robustness and consistency tests discussed in the End Matter show that the P70 mask  yields a slightly tighter upper limit $\ymono = \left(1.2 \pm 2.0\right) \times 10^{-6}$ ($\ymono<5.2\times 10^{-6}$) consistent with those obtained on P60, P40, and P20. Therefore, we adopt this as our final estimate.

\section{Cosmological and astrophysical implications}\label{sec:feedback}

A joint measurement of $\ymono$ and the monopole of its first-order relativistic corrections, $\tmono$, can constrain combinations of energy injections in the gas (commonly referred to as baryonic feedback) by active galactic nuclei (AGN) and supernovae (SN) in low-mass haloes and groups ($M\lesssim 10^{14}\msun$) \cite[][hereafter \citetalias{thiele2022}]{thiele2022} with sub-percent precision. These systems are among the most affected by feedback and also the ones to which cosmological constraints from galaxy lensing are the most sensitive to \cite{halo_model_review}. Spectral distortions measurements thus provide information on feedback complementary to that in the tSZ angular power spectrum or higher-order statistics, which are dominated by galaxy clusters contributions ($10^{14} \, \msun \lesssim M \lesssim 10^{15} \, \msun$). 
 
To interpret our $\ymono$ measurement in this context, we used the CAMELS simulations suite \cite{camels,camels-astrid}, which spans a wide range of galaxy formation models. CAMELS varies 4 feedback parameters describing SN ($A_{SNi}$) and AGN-generated ($A_{AGNi}$) winds (where $i=1$ components controls the amplitude of energy injections, and the $i = 2$ the speed dependence of the outflows) alongside $\sigma_8$ and $\Omega_m$ cosmological parameters over a latin hypercube (LH) for several hydrodynamical models (SIMBA \cite{simba}, Astrid \cite{astrid2022a,astrid2022b}, IllustrisTNG \cite{illustristng2017,illustristng2018}). We compute $\ymono$ and $\tmono$ predictions for all 6144 CAMELS simulations integrating directly the electron pressure, and accounting for cosmic variance, volume effects, and the halo mass range limitations of CAMELS as in~\citetalias{thiele2022}\footnote{These corrections are computed using a halo model-based approach \cite{hill2015}.}. We also include predictions from additional CAMELS suites: Astrid LH, SB7 (additionally varying $\Omega_b$ on a Sobol sequence, SB) and IllustrisTNG SB28 (varying $\Omega_b, h, n_s$ and all IllustrisTNG model parameters).

\begin{figure*}[!htbp]
\centering
\includegraphics[width=\textwidth]{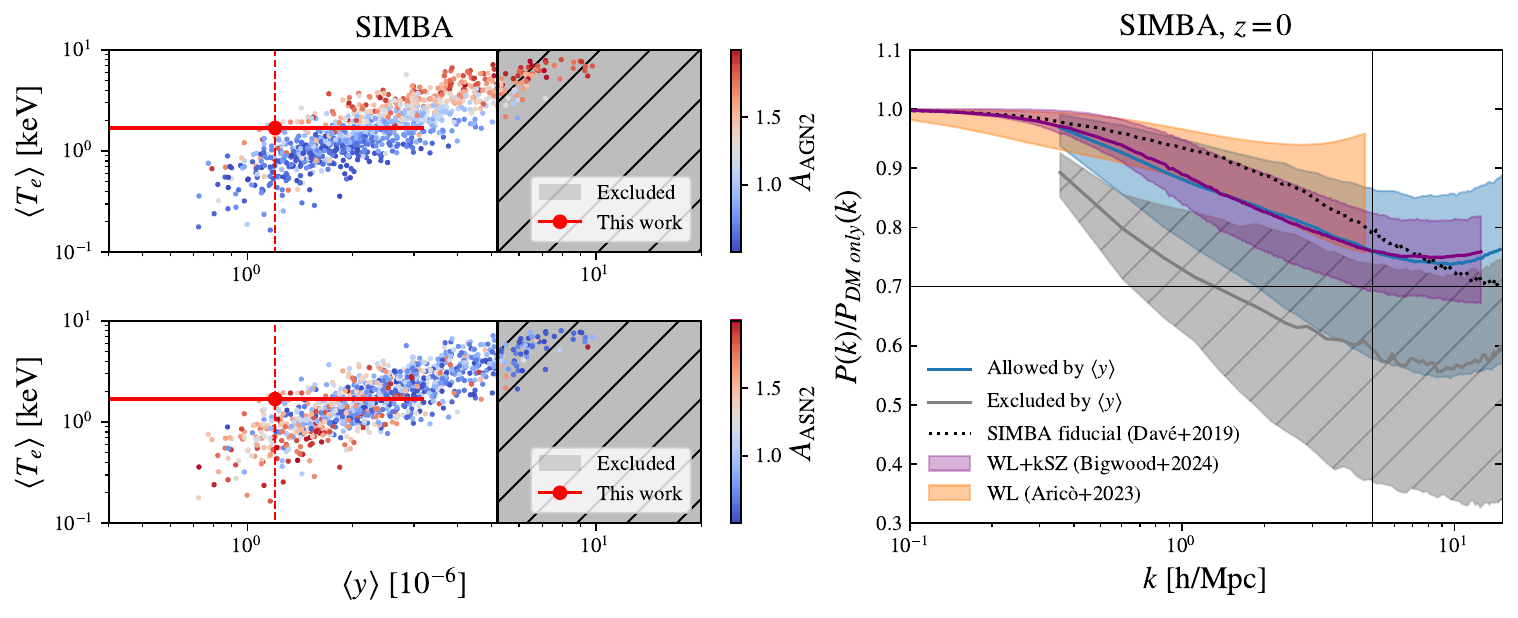}
\caption{Left: CAMELS SIMBA predictions of $\tmono$, $\ymono$ color coded by values of $A_{AGN2}$ (top) and $A_{SN2}$ (bottom)  compared to our $\ymono$ measurement (red). As $\tmono$ is unconstrained by the data, we fix it to the median of the predictions. The excluded area at 95\% C.L.~is shown in grey. 
 Right: baryon-induced matter power spectrum suppression measured in the CAMELS SIMBA LH suite with allowed values constrained from DES cosmic shear alone (orange) and with kSZ data (purple). The median of the excluded (allowed) models in the left panels is shown in grey (blue) and the corresponding shaded area show their 16th and 84th percentiles. The suppression of the fiducial SIMBA simulation appears in black \cite{simba}.}
\label{fig:implications}
\end{figure*}

Among these, SIMBA is the model with the strongest feedback and covering the widest prediction range, reflecting thus more representatively current large uncertainties in feedback physics. We therefore focus on this model, showing its $\ymono$, $\tmono$ predictions alongside our $\ymono$ measurement in Fig.~\ref{fig:implications} after rescaling them to the reference CAMELS cosmology using the multi-dimensional interpolator of \citetalias{thiele2022}. 
The spread of the predictions thus arises purely from variations in feedback processes. Even without direct $\tmono$ constraints, our $\ymono$ measurement alone rules out $\sim10\%$ of SIMBA models. The FIRAS sensitivity and the absence of a $\tmono$ measurement prevents a selection of a preferred feedback model, due to parameter degeneracies, but we can exclude parameter combinations with low $A_{SN2}$ and high $A_{AGN2}$, while $A_{SN1}$ and $A_{AGN1}$ remain unconstrained. Results for other suites are discussed in the Supplemental Material.

A key consequence of feedback for cosmological studies is the expulsion of baryons from gravitationally bound dark matter halos, which suppresses structure formation at scales $k>1h/{\rm Mpc}$ and introduces degeneracies with cosmological parameters \cite{salcido2025}. This strongly affects the cosmological interpretation of galaxy lensing data. Recent works suggests that tensions between the matter fluctuations amplitude $S_8=\sigma_8\sqrt{\Omega_m/0.3}$  inferred from galaxy lensing and CMB measurements could be reconciled by assuming stronger feedback than what is often assumed in state-of-the-art hydrodynamical simulations. This has been supported by joint galaxy lensing and tSZ or kSZ analyses \cite{arico2023,hadzhiyska2024,bigwood2024,pandey2025}. The SIMBA feedback model could be strong enough to alleviate these tensions \cite{schaller2025b}.
 We therefore examined the suppression of the 3D matter power spectrum ($P(k)/P(k)^{\rm DM,only}$) caused by feedback relative to the case where the gravitational collapse of dark matter is the only acting force in structure formation, focusing on models excluded by our $\ymono$ measurement. We neglected the residual cosmological parameters dependence on this quantity as it is mild \cite{schaller2025}. As shown in Fig.~\ref{fig:implications}, suppression levels in models excluded at high significance by our $\ymono$ measurement are also excluded by cosmic shear and kSZ data, though degeneracies in the feedback sector can also allow for the opposite to happen. However, part of the baryonic suppression range favored by current cosmic shear and kSZ measurements may already be inconsistent with our $\ymono$ constraints. While a full joint analysis of cosmic shear, kSZ and $\ymono$ data is beyond the scope of this Letter, and these last findings are to some extent model-dependent, our results suggest that our new $\ymono$ measurement is now precise enough to directly inform on feedback models and could serve as an additional, complementary probe alongside other observables. 

\section{Conclusions}
The threefold improvement in precision over the original FIRAS results achieved by our analysis approaches the sensitivity needed for a direct detection of  $\ymono$ from structure formation, and makes $\ymono$ informative for baryonic feedback studies. 
While our results are robust to foreground contamination, uncertainties in foreground knowledge and modeling remain an important concern for future experiments. Assuming BISOU- and FOSSIL-like sensitivities of $\sim 10^{-4}$ and $\sim 10^{-6}$ MJy/sr respectively \cite{bisou2024,coulon2024}, a sky model from \cite{Abitbol:2017vwa} (16 free parameters total including CMB, CIB, Galactic dust, synchrotron with 10\% Gaussian priors on its amplitude and spectral index, free-free, spinning dust (AME), and integrated CO), and a frequency range of 30–2010 GHz in $\Delta \nu = 15$ GHz bins, we obtain Fisher errors $\sigma_{\ymono}\sim 10^{-6}$ ($1.6 \times 10^{-8}$) for BISOU (FOSSIL) adopting a \textit{frequency–monopole} component separation method. The \textit{pixel-by-pixel} foreground cleaning adopted here typically outperforms the \textit{frequency–monopole} by 3-5 times \cite{sabyr2025}, implying an achievable detection significance $\gg 5$ for BISOU and $\gg 100$ for FOSSIL for $\ymono \sim 2\times 10^{-6}$, and a measurement with signal-to-noise $\gg 10$ of $\tmono=1.25$keV for FOSSIL. These would establish $y$ distortions as a precise cosmological and astrophysical probe even in presence of complex foreground emissions.

\ifwordcount
\else
\section*{Acknowledgments}
We thank Luca Pagano, Shy Genel and Francisco Villaescusa-Navarro for useful discussions and support throughout this work, and Leah Bigwood and Giovanni Aric\`o for providing the data used in Fig.~\ref{fig:implications} and discussions. GF acknowledges the support of the European Research Council under the Marie Sk\l{}odowska Curie actions through the Individual Global Fellowship No.~892401 PiCOGAMBAS and of the Simons Foundation, where this work was started. He also acknowledges the support of the European Union’s Horizon 2020 research and innovation program (Grant agreement No. 851274) and of the STFC Ernest Rutherford fellowship during the final stages of this work.  AS and JCH acknowledge support from NASA grant 80NSSC23K0463 (ADAP).  JCH also acknowledges support from NASA grant 80NSSC22K0721 (ATP), the Sloan Foundation, and the Simons Foundation. LT is supported by JSPS KAKENHI Grant 24K22878. Part of this research used computational resources of the National Energy Research Scientific Computing Center (NERSC), a U.S. Department of Energy Office of Science User Facility operated under Contract No. DE-AC02-05CH11231, and of the Simons Foundation.  Some of the results in this letter have been derived using the \texttt{healpy}/\textsc{Healpix} \cite{Zonca2019,healpix}, getdist \cite{getdist}, NumPy \cite{2020NumPy-Array}, SciPy \cite{2020SciPy-NMeth} and Matplotlib libraries \cite{Hunter:2007}.

\section*{Author Contributions}
GF and FB carried out the analysis for the $\ymono$ measurements and conceptualized the work with DS. AS and JCH contributed to the definition of the component separation algorithm and robustness tests of the results. LT provided numerical tools to obtain the simulation-based predictions for $\ymono$ and $\tmono$ in CAMELS. GF and CL carried out the analysis on feedback parameters. All authors contributed to the interpretation of the results and the writing of the manuscript.
\fi

\ifwordcount
\else
\bibliographystyle{apsrev4-2}
\bibliography{specdist_firas_y}
\fi

%appendices / end matter
\ifwordcount 
\else
% !TEX root = main_prl.tex

\clearpage

\section*{$\ymono$ robustness and consistency tests}\label{appendix:consistency}
\begin{table*}[!htbp]
 \begin{tabular}{lSSSSSS}
 \toprule\toprule
 \multicolumn{6}{c}{$\langle y\rangle\ [\times 10^{-6}]$}                    \\ 
 \midrule
 Foreground model     & \multicolumn{1}{c}{\phantom{-0.8$\pm$}P80}& \multicolumn{1}{c}{\phantom{-0.8$\pm$}P70}\  &\multicolumn{1}{c}{\phantom{-0.8$\pm$}P60}  &\multicolumn{1}{c}{\phantom{-0.8$\pm$}P40} &\multicolumn{1}{c}{\phantom{-0.8$\pm$}P20}&\multicolumn{1}{c}{$<95\%$ C.L.} \\
 \midrule
$A_d + \beta_{d}$	                         &-0.8\pm1.6&0.1\pm 1.7& 2.0\pm 1.8$^{\dagger}$& 4.2\pm 2.1& 6.3\pm 2.9&5.5\\
$A_d +A_{CIB}^{Abitbol}$      	        &0.3 \pm 2.9&0.3\pm 3.0& 2.7\pm 3.2$^{\dagger}$& 0.9\pm 3.9& 2.9\pm 5.6&9.2\\
$A_d +A_{CIB}^{Fixsen}$		        &-0.4\pm 3.0&-0.8\pm 3.1&0.7\pm 3.4$^{\dagger}$&-0.4\pm 4.0&-0.8\pm 5.7&7.5\\
$A_d +A_{CIB}^{Gispert}$ 	        &4.4 \pm 2.5&3.8\pm 2.6&5.2\pm 2.8$^{\dagger}$&4.8\pm 3.3&5.1\pm 4.8&10.9\\
$A_d + \beta_d+A_{CIB}^{Abitbol}$  &-2.7\pm 3.6&-1.1\pm 3.8&1.4\pm 4.2$^{\dagger}$&-3.7\pm 5.5&-0.5\pm 8.9&9.9\\
$A_d + \beta_d+A_{CIB}^{Fixsen}$  &0.7 \pm 3.1&-2.5\pm 3.3&1.1\pm 3.7$^{\dagger}$&-3.5\pm 4.8&-3.3\pm 8.0&8.4\\
$A_d + \beta_d+A_{CIB}^{Gispert}$ &2.0 \pm 4.4&0.4\pm 4.8&0.1\pm 5.3$^{\dagger}$&-4.9\pm 7.0&0.1\pm 11.3&10.8\\
$A_d + \beta_d+T_d$		        &0.8 \pm 1.9&1.2\pm 2.0$^{\dagger}$&1.7\pm 2.1&0.7\pm 2.8&-6.9\pm 4.7&5.2\\
\bottomrule\bottomrule
\end{tabular}
\caption{Summary of all the measurements of $\ymono$ obtained for different foreground models and Galactic masks. We assumed a $\nu_{max}=600$ GHz. $^{\dagger}$ indicates the reference value for which we computed the 95\% C.L. upper limit.}
\label{tab:ymono-allmasks}
\end{table*}
Table \ref{tab:ymono-allmasks} shows that for our most sensitive $\ymono$ measurement ($A_d + \beta_d$ model), the results obtained from the P70, P80, P90 masks are in $3\sigma$ tension (using Eq.~\ref{eq:yshift}) with those obtained from P60, P40, and P20. For the $A_d + \beta_d + T_d$ model, $\ymono$ estimates remain consistent within $\sim 2.3\sigma$ up to P80 (i.e. only slightly above the acceptability threshold), while P90 shows $3\sigma$ tension with P80 and P70. The tightest upper limit including the largest sky fraction and consistent with more conservative Galactic masks, $\ymono < 4.6 \times 10^{-6}$, is obtained with the $A_d + \beta_d + T_d$ model on the P80 mask. Since we  recover an unbiased $\ymono$ on mock data for P80 and $A_d + \beta_d + T_d$, we test further whether the P80 region contains residual foreground contamination that biases $\ymono$ low. To this end, we defined three sets of approximately disjoint masks based on differences between the \planck Galactic masks, since they are nested:
\begin{itemize}
\item $P60_{set}$ = $\{\rm{P90-P20, P80-P20, P60}\}$
\item $P40_{set}$ = $\{\rm{P80-P40, P60-P20, P40}\}$
\item $P20_{set}$ = $\{\rm P90-P70, P80-P60, P60-P40, P40-P20, P20 \}$
\end{itemize}
Elements of each set cover a similar total sky area. We computed $\ymono$ for every mask in each set and tested their consistency using Eq.~\ref{eq:yshift}. If consistent, we combined them through an inverse-covariance weighted average (to account for correlation induced by any overlapping sky area). Since pixel correlations in the component-separated $y$ maps are small \cite{sabyr2025}, these averages should achieve the same statistical power as a single $\ymono$ estimate over an equivalent  sky fraction (e.g., averaging estimates of the $P20_{set}$ corresponds to the sensitivity achievable on $\sim 40\%$ of the sky). This test evaluates whether the additional sky area in P80 (or a subset of it) introduces anomalous shifts in $\ymono$. The results are summarized in Table \ref{tab:robustness}. 
We found no significant tension among $\ymono$ estimates from the $P40_{set}$ and $P60_{set}$ masks. In the $P20_{set}$, however, the value from ${\rm P90-P70}$ agrees with ${\rm P80-P60}$ but is in $3\sigma$ tension with estimates from masks extending farther from the Galactic plane. ${\rm P80-P60}$ itself is consistent with ${\rm P60-P40, P40-P20}$, and ${\rm P20}$. Because $\ymono = (-14.4 \pm 6.0) \times 10^{-6}$ on ${\rm P90-P70}$, and unusually large negative values are often linked to foreground residuals \cite{sabyr2025}, we investigated this further.
We created a new Galactic mask retaining 50\% of the sky (P50) following the same procedure of \planck \cite{planck18_likelihood}, and defined a new $P10_{set} = \{\rm P90-P80, P80-P70, P70-P60, P60-P50, P50-P40\}$, composed of disjoint masks covering 10\% of the sky to isolate the additional areas introduced when including more sky area going from P50 to P90 masks. Within $P10_{set}$, $\ymono$ from ${\rm P80-P70}$ is in $3\sigma$ tension with ${\rm P50-P40}$, and ${\rm P90-P80}$ is inconsistent with nearly all other subsets.
As such, because the added area between P70 and P80 shows partial inconsistency with cleaner, higher Galactic latitude regions, and it overlaps with parts of P90 known to suffer from foreground contamination, we adopt the P70 mask as our most conservative upper limit. However, since the area added between P60 and P80 is consistent with comparable higher-latitude subsets, we include it in our consistency tests.
Excluding ${\rm P90-P70}$ from the $P20_{set}$ and averaging the remaining regions yields a $\ymono$ value consistent with that from P40, as expected given the comparable statistical precision in the absence of systematics. Likewise, the averages from $P40_{set}$ and $P60_{set}$ agree with the results from P60 and P80, respectively. All these estimates are mutually consistent, indicating only a minor, if any, impact from foreground residuals.

As an additional robustness test, we checked the stability of our results by extending the frequency range to $\nu_{max} = 1$THz. Although higher frequencies can improve foreground separation, \citetalias{sabyr2025} showed that the larger systematic errors of high-frequency FIRAS data and increased sky complexity can make such analyses more sensitive to foreground residuals.
Using $\nu_{max} = 1$ THz on mock data, we found that models including $A_d + \beta_d + A_{CIB}$ yield unbiased $\ymono$ estimates up to a P60 mask, whereas $A_d + \beta_d + T_d$ must be restricted to smaller sky fractions. Other methods produced biased $\ymono$ values, including $A_d + \beta_d + A^{Gispert}_{CIB}$. Since the Gispert SED differs the most from the baseline SED adopted in the mock data, discrepancies among $A_d + \beta_d + A_{CIB}$ models with different CIB SEDs likely indicate sensitivity to CIB residuals.
We observed a similar trend in the data but report in Table~\ref{tab:robustness} the $\ymono$ values from $A_d + \beta_d + A^{Abitbol}_{CIB}$ and $A_d + \beta_d + A^{Fixsen}_{CIB}$, which remain consistent with our baseline results. We estimate a systematic uncertainty of $\sigma^{syst}_{\ymono} \approx 0.4 \times 10^{-6}$ in $\ymono$ from the standard deviation of these results, and including this uncertainty does not alter our final upper limit within rounding errors.

% robustness tests
\begin{table}[htbp]
\begin{tabular}{lSS}
\toprule\toprule
Foreground model & \multicolumn{1}{c}{\phantom{$2.0 \pm$}$\ymono [10^{-6}]$} & \multicolumn{1}{c}{$<95\%\ {\rm C.L.}$}\\
\midrule
$A_d+\beta_d+T_d$ P60 &1.7 \pm 2.1&5.9\\
$A_d+\beta_d+T_d$ P40 &0.7\pm 2.8&6.3\\
$A_d+\beta_d+T_d$ P80 &0.8\pm 1.9&4.7\\
$A_d+\beta_d+T_d$ $P60_{set}$ &0.9 \pm 1.9&4.8\\
$A_d+\beta_d+T_d$ $P40_{set}$ &1.3 \pm 2.1&5.5\\
$A_d+\beta_d+T_d$ $P20_{set}$ &2.1 \pm 2.2&6.4\\
\midrule 
$A_d+\beta_d+A^{Abitbol}_{CIB}$ $\nu_{max}=1$THz &2.0 \pm 1.8&5.5\\
$A_d+\beta_d+A^{Fixsen}_{CIB}$ $\nu_{max}=1$THz &1.6 \pm 1.9&5.4\\
\bottomrule\bottomrule
\end{tabular}
\caption{Stability of $\ymono$ measurement to analysis variations.}
\label{tab:robustness}
\end{table}

\section*{Robustness of feedback constraints}
In hydrodynamical simulations, feedback strength is often linked to the baryon fraction  $f_b=\Omega_b/\Omega_m$. Since the SIMBA LH suite in CAMELS keeps $\Omega_b$ fixed while varying $\Omega_m$, models excluded by $\ymono$ could, in principle, correspond to low-$\Omega_m$ regions and thus anomalously high-$\Omega_b$, therefore artificially amplifying feedback effects. 
However, as shown in Fig.~\ref{fig:all_fb}, this is not the case: the median $f_b$ of the excluded SIMBA models, though slightly higher, remains consistent within $1\sigma$ of the \planck-like CAMELS cosmology. Errors on the median were estimated via bootstrap resampling of $f_b$ values from the full LH suite. In contrast, weaker feedback models, such as those in Astrid and IllustrisTNG, are ruled out by our $\ymono$ measurement only for cases with high $f_b$, i.e. where feedback is artificially enhanced and rescaling of cosmology dependence becomes less accurate. Constraints from these models are therefore less robust. For Astrid, we could not apply the specific correction accounting for missing massive halos in the limited CAMELS volume, as the large parent volume of Astrid was not publicly available at all redshifts, leading to biased-low $\ymono$ values. 
Despite this limitation, Astrid, like SIMBA, allows us to exclude parts of the parameter space explored in CAMELS, particularly models with low $A_{SN2}$ and $A_{AGN2}$, corresponding to weak supernova feedback and low AGN jet energy injection (see \cite{camels-astrid} for a more detailed discussion on the meaning of these parameters). By contrast, no specific region is excluded for the IllustrisTNG suite. We show all these results in the Supplemental material. We further tested if the models ruled out in Fig.~\ref{fig:implications} could identify a specific combination of feedback parameters using the neural posterior estimation method of \cite{ltu-ili} and a simulation based inference approach, in order to have a better treatment of internal feedback parameter degeneracies. While we found that $A_{AGN2}$ could be marginally constrained, degeneracies remain the dominant effect and prevent a robust direct measurement. Nonetheless, in Fig.~\ref{fig:simba-hist} we show how our $\ymono$ measurement rules out specific regions of the parameter space. The preferred median values of $A_{SN2}$ and $A_{AGN2}$ in the excluded models cannot in fact be explained with a random combination of feedback model parameters. 

\begin{figure}[!htbp]
\includegraphics[width=\columnwidth]{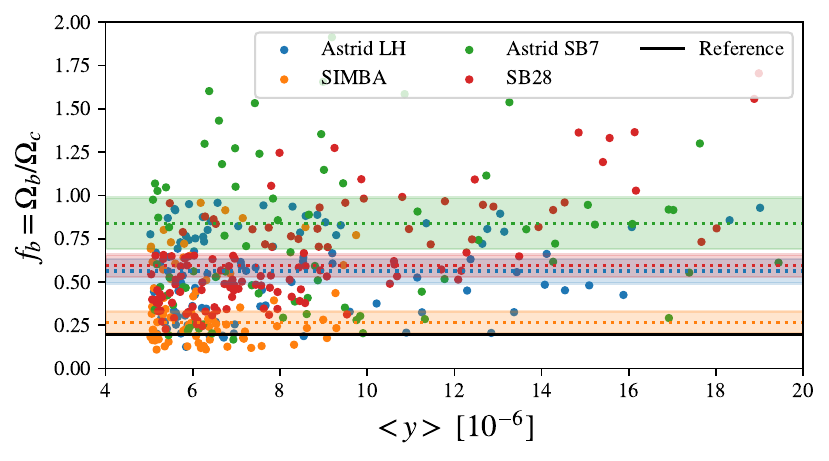}
\caption{Baryon fraction $f_b$ of each of the CAMELS simulations excluded by our $\ymono$ measurement. The dotted line and shaded area show the median $f_b$ and its bootstrap error.}
\label{fig:all_fb}
\end{figure}

\begin{figure}[!htbp]
\includegraphics[width=.8\columnwidth]{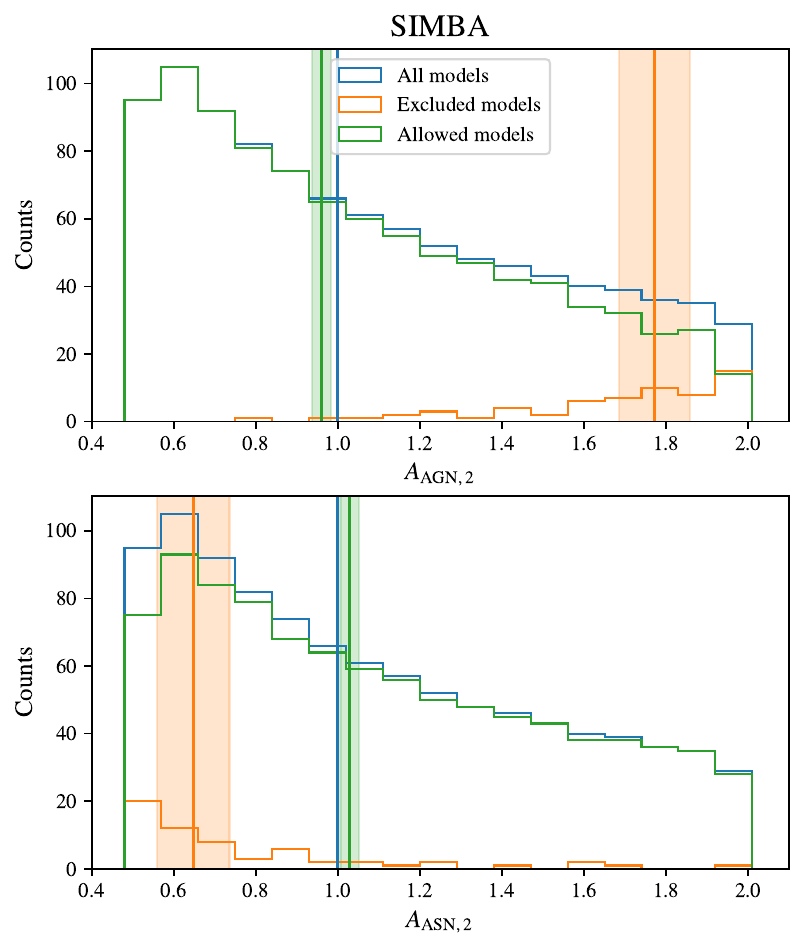}
\caption{Histogram of the values of $A_{AGN2}$ (top) and  $A_{SN2}$ (bottom) for the models allowed and excluded by $\ymono$ in the CAMELS SIMBA suite. The corresponding median and bootstrap error are shown as a solid line and shaded area.}
\label{fig:simba-hist}
\end{figure}

% !TEX root = main_prl.tex

%
\clearpage
\onecolumngrid
\section*{Supplemental material}
\setcounter{figure}{0}
\setcounter{table}{0}

\begin{table}[htbp]
\begin{tabular}{lSS}
\toprule\toprule
Foreground model & \multicolumn{1}{c}{\phantom{$2.0 \pm$}$\ymono [10^{-6}]$} & \multicolumn{1}{c}{$<95\%\ {\rm C.L.}$}\\
\midrule
$A_d+\beta_d$ & 1.2 \pm  1.7&4.6\\
$A_d+A_{CIB}^{Abitbol} $ & -0.8 \pm 3.1&5.4\\
$A_d+A_{CIB}^{Fixsen}  $ &-2.4 \pm 3.2&4.0\\
$A_d+A_{CIB}^{Gispert}$ &2.1 \pm 2.7&7.5\\
$A_d+\beta_d+A_{CIB}^{Abitbol}$ &2.9 \pm 3.3&9.5\\
$A_d+\beta_d+A_{CIB}^{Fixsen}$ &1.3 \pm 3.2&7.7\\
$A_d+\beta_d+A_{CIB}^{Gispert}$ &3.0\pm4.6&12.2\\
$A_d+\beta_d+T_d$ &0.7\pm2.0&4.7\\
\midrule 
$A_d+\beta_d+T_d$, P80 &0.7 \pm 1.8&4.3\\
$A_d+\beta_d+A_{CIB}^{Abitbol}$, $\nu_{max}=1$THz&-2.6 \pm 1.5&0.4\\
$A_d+\beta_d+A_{CIB}^{Fixsen}$, $\nu_{max}=1$THz &-2.2 \pm 1.6&1.0\\
\bottomrule\bottomrule
\end{tabular}
\caption{Summary of $\ymono$ measurements obtained for the reference setups of Table \ref{tab:ymono-allmasks} and Table \ref{tab:robustness} in the main text on mock data. Further details on the sky modeling adopted in the mocks are described in \citetalias{sabyr2025} and in the main text.}
\label{tab:mocks}
\end{table}

\begin{figure*}[!htbp]
\includegraphics[width=\textwidth]{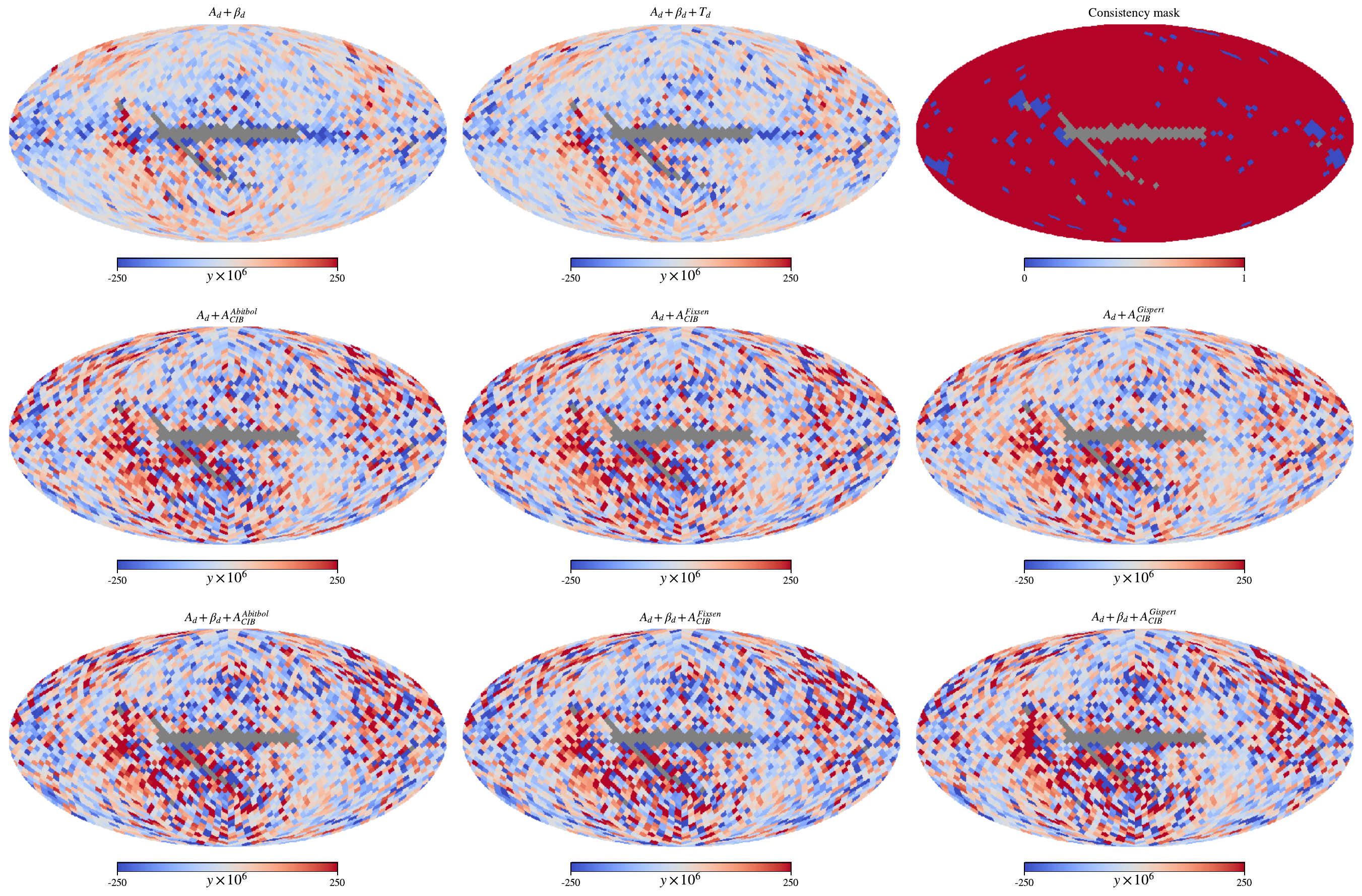}
\caption{$y$ maps obtained with different component separation methods investigated in this work. The top right panel shows the consistency mask that identifies pixels for which different component separation methods give consistent estimates (red). Pixels removed by the FIRAS destriper mask appear in grey.}
\label{fig:all_y_maps}
\end{figure*}

\begin{figure*}[!htbp]
\includegraphics[width=\textwidth]{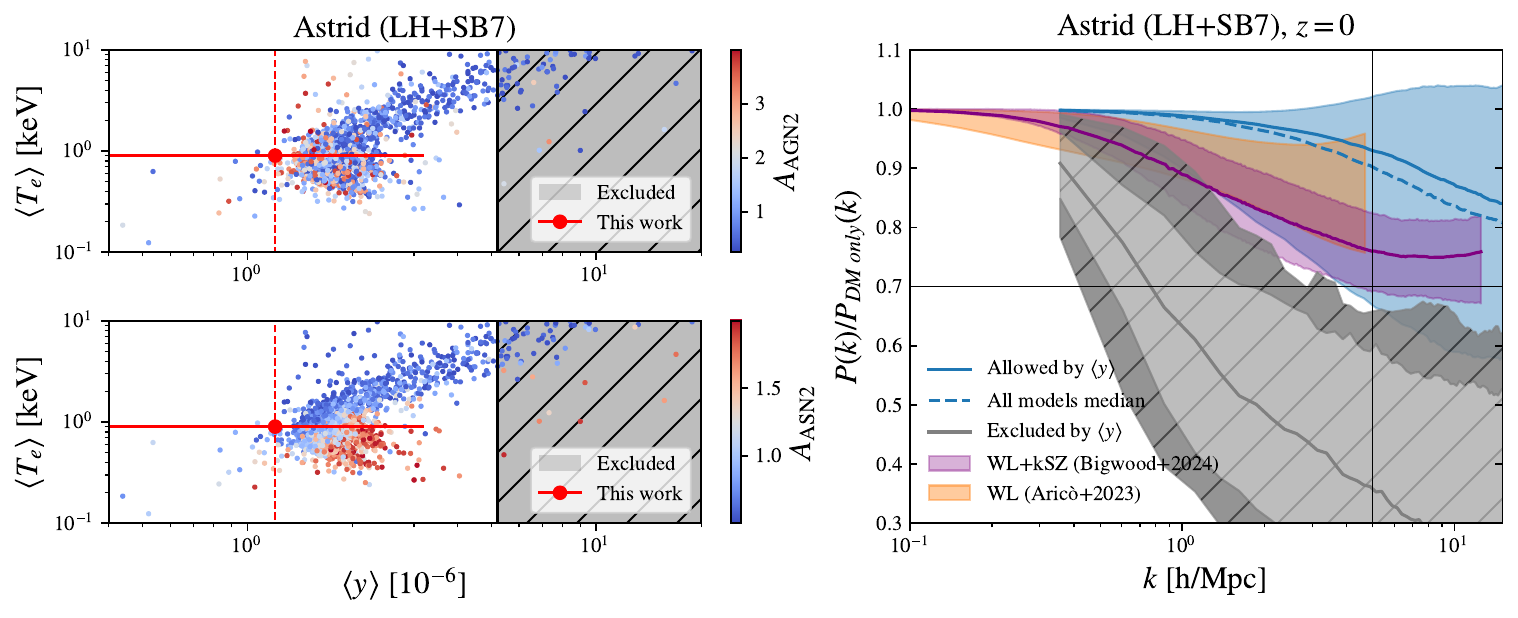}
\caption{As Fig.~\ref{fig:implications} in the main text for Astrid LH and SB7 suites. The darker grey shaded areas show the 95th percentile.}
\label{fig:all_pk_constraints}
\end{figure*}
\begin{figure}[!htbp]
\includegraphics[width=.89\columnwidth]{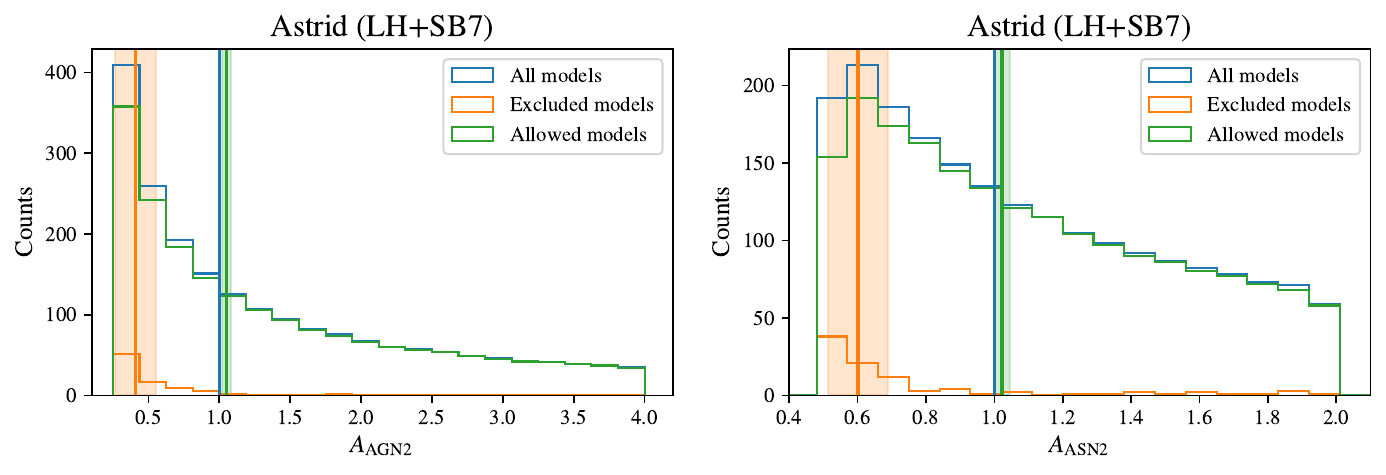}
\caption{Same as Fig.~\ref{fig:simba-hist} in the main text for the CAMELS Astrid LH and SB7 suites.}
\label{fig:astrid-hist}
\end{figure}

\begin{figure*}[!htbp]
\includegraphics[width=\textwidth]{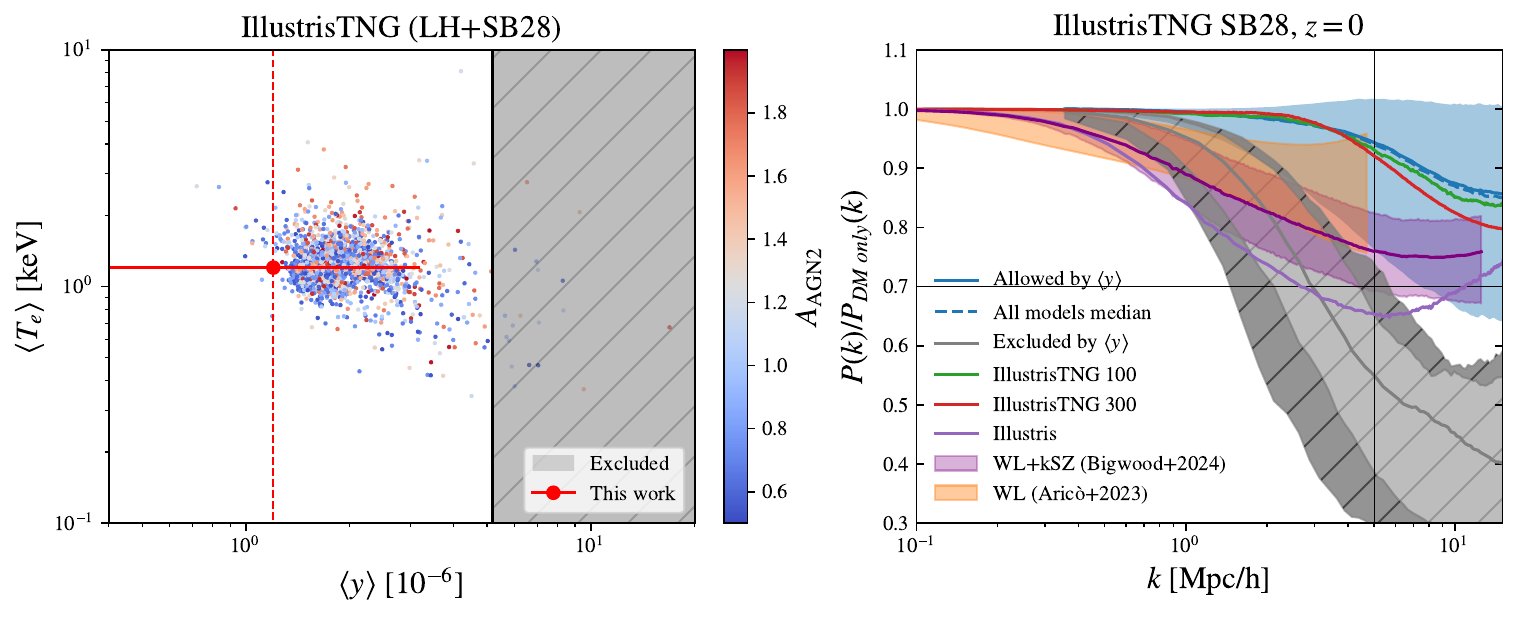}
\caption{As Fig.~\ref{fig:all_pk_constraints} above for IllustrisTNG LH and SB28 suites. We show the fiducial suppression of the Illustris \cite{illustris}, IllustrisTNG 100 and IllustrisTNG 300 \cite{illustristng2018,illustristng2018b} simulations in solid lilac, green and red respectively.}
\label{fig:all_pk_constraints_tng}
\end{figure*}

\fi

\end{document}